# Energy Aware Node Selection for Cluster-based Data Accuracy Estimation in Wireless Sensor Networks

Jyotirmoy Karjee ,H.S Jamadagni
Centre for Electronics Design and Technology
Indian Institute of Science
Bangalore, India
kjyotirmoy@cedt.iisc.ernet.in, hsjam@cedt.iisc.ernet.in

***Abstract*****—**Objective : The main objective of this paper is to reduce the number of sensor nodes by estimating a trade off between data accuracy and energy consumption for selecting nodes in probabilistic approach in distributed networks. *Design Procedure/Approach:* Observed data are highly correlated among sensor nodes in the spatial domain due to deployment of high density of sensor nodes. These sensor nodes form non-overlapping distributed clusters due to high data correlation among them. We develop a probabilistic model for each distributed cluster to perform data accuracy and energy consumption model in the network. Finally we find a trade off between data accuracy and energy consumption model to select an optimal number of sensor nodes in each distributed cluster. We also compare the performance for our data accuracy estimation model with information accuracy model for each distributed cluster in the network. *Practical Implementation:* Measuring temperature in physical environment and measuring moisture content in agricultural field. *Inventive /Novel Idea:* Optimal node selection in probabilistic approach using the trade of between data accuracy and energy consumption in cluster-based distributed network.

***Keywords - Spatial correlation; distributed clusters;data accuracy;energy consumption;tradeoff ; wireless sensor networks***

## I. INTRODUCTION

Recent improvements in wireless communication have made a drastic development over wireless sensor networks and wireless embedded systems. Because of the reliable cost, ease of development, small size, wireless sensors are used in many applications such as collecting data of any event like temperature, humidity, seismic event. Wireless sensor is a small processing device called node which captures raw data from the physical environment[1], process it, communicate the raw data wirelessly among other nodes and finally transmit the collected raw data to the sink node. The raw data collected by the sensor nodes are generally spatially correlated [2] among other sensor nodes in the physical environment. As the density of the sensor nodes increases the spatially proximal sensor observation among the sensor nodes are highly correlated [3] in the sensor field. The sensor nodes form distributed clusters [4] in the sensor field due to high data correlation among the sensor nodes. Formation of distributed clusters minimizes data collection cost [5]. LEACH [6] gives a clear idea about the formation of distributed dynamic clusters in the sensor field according to priori probability. Each dynamic distributed clusters have a cluster head (CH) [7,28] node which collects the observed data and transmits it to the sink node. Like LEACH, SEP [8] forms distributed clusters in the heterogeneous networks. According to literature [9], sensor nodes forms spatial correlation of data among them where the data correlation shows Gaussian distribution with zero mean. In literature [10] a grid based clustering method is proposed. But these types of theoretical clustering methods rarely happen in original scenario. In literature [11] a disk shaped circular clustering algorithm is proposed where sensor nodes groups into disjoint set with a designated CH node. This type of disk shape cluster doesn't form in real case. In practical scenario, clusters are irregular in shape and size. In literature [4], authors proposed a distributed clustering algorithm with different shape and size based upon shortest distance among associated sensor nodes and CH node in the sensor field. In literature [16], authors proposed a new non-overlapping distributed clustering algorithm with different sizes based upon spatially correlated data among sensor nodes in the spatial domain. We adopt the clustering algorithm [16], since this non overlapping clustering algorithm has more practical sense of implementation in the spatial domain. The size of each distributed cluster in this clustering algorithm is based upon a threshold value given in data correlation model [4,16] in the sensor field.

According to literature [12,13,14,20], authors proposed information accuracy (distortion function) model where sink node can estimate the information accuracy for observed data sensed by all the sensor nodes. These proposed information accuracy model follows the single hop communication where the observed data are directly transmitted to the sink node. But in literature [4,15,16],authors proposed two hop communication where the observed data are transmitted to the sink node via CH node for each distributed cluster in the network. Generally observed data are directly transmitted to the CH node for each distributed cluster for aggregation [19,26,27] without verifying the data accuracy. This increases the redundant data in the CH node and also increases the transmission overhead in the network. Hence we verify the data accuracy using Minimum Mean Square Error (MMSE) estimator [25] before data aggregation at CH node for each distributed cluster in the network. Sometimes due to extreme external physical environment such as heavy rain fall, snow fall etc, some of the sensor nodes get malicious [18]. In such situation sensor nodes can sense and read inaccurate data in the cluster. In accurate data transmitted by malicious node may cause incorrect data aggregation at the CH node which also increases the data redundancy in the network. Hence it is very essential to estimate and verify the data before aggregation at the CH node for each distributed clusters in the sensor network.

In this paper, we use the proposed data accuracy model which uses MMSE estimator given in the literature [16 ] and compare with information accuracy [12] model .According to literature [12] , MMSE estimation is done at each individual sensor node for the observed data before transmitting the estimated data at the CH node in each distributed cluster in the network. Once the estimated data is received at the CH node transmitted by all the sensor nodes in the cluster, we average it and finally transmit it to the sink node. This increases the communication overhead in the network .But in data accuracy model, we calculate the MMSE estimation only at the CH node for all the sensor nodes in the cluster which increases the data accuracy for each distributed cluster and reduces the communication overhead in the network .

In literature [29], the author proposes a node selection scheme where the network reduces energy consumption by identifying redundant nodes with respect to sensing coverage. In literature [17], authors proposes selection of nodes using GB (Grid based) node choice algorithm which save energy in the network. Similarly in literature [20], authors minimizes energy consumption with optimizing the maximum information accuracy for a cluster based WSN by using an information accuracy aware jointly sensing nodes selection algorithm. Hence these work deals with reducing the number of sensor nodes in the network with respect to sensing coverage and also finding an information accuracy aware jointly sensing nodes selection algorithm based on maximum information accuracy with minimum energy consumption. But in this paper we propose a novel probabilistic model to select an optimal sensor node in each distributed clusters by finding a trade off between data accuracy and energy consumption. We also find the probability by which the sensor nodes are in active mode and sleep mode in each distributed cluster subjected to data accuracy and energy consumption. Finally this leads to select an optimal sensor node which reduces the energy consumption and increases the lifetime of the network.

The main objective of this paper is to reduce the number of sensor nodes per cluster in the distributed network by finding a trade off between data accuracy and energy consumption in probabilistic approach. Hence the flow of this paper is given as follows. In section II , we summarize our mathematical model proposed in literature [16] to extend our new work which is discuss in Section III.Initially we deploy all the sensor nodes in the sensor field. Sensor nodes form data correlation [16] among observed data sensed by them discussed in section II A. Since the observed data are highly correlated among sensor nodes, sensor nodes form non-overlapping distributed clusters among them in the sensor field explained in section II B. Once the sensor nodes form non-overlapping disjoint groups of distributed clusters, each distributed cluster perform data accuracy at the respective CH node discussed in Section II C. In section III, we propose a probabilistic model for selecting some of the sensor nodes to be switched on in each distributed cluster for transmitting the data to the sink node. In Section III A –B, we find by what probability the sensor nodes are switch on and switch off in each distributed cluster can be controlled subjected to data accuracy and energy consumption respectively in the whole network. This leads to save energy in the whole network by controlling some of the sensor nodes to be in sleep mode for each distributed cluster. It leads to find a trade off between data accuracy and energy consumption for each distributed clusters in the network discussed in section-III C. This trade off reduces node deployment cost and energy cost per cluster in the distributed network. Finally this trade off between data accuracy and energy consumption for each distributed cluster is implemented to reduce the number of sensor nodes for the whole network.

## II. Cluster-based Data Accuracy Estimation in Spatial Domain

Initially sensor nodes are randomly deployed in the sensor field. If we keep on deploying the sensor nodes, the sensor node density increases and the spatial correlation of the observed data among the sensor nodes also increases in the sensor field. Since the spatial correlation of data increases among the sensor nodes, we get more accurate observed data from the sensor nodes. But as the number of sensor nodes increases, the node deployment cost and the energy consumption in the network also increases. Hence we have to reduce the node deployment cost by reducing the number of sensor nodes such that we also achieve better accurate observed data from the sensor nodes with less energy consumption in the network.

In this section, we discuss a mathematical foundation for the spatial correlation of data among sensor nodes to form non-overlapping distributed clusters in the sensor field. Once the non-overlapping distributed clusters are constructed in the sensor field, we investigate to what degree the accurate data are extracted by sensor nodes in each distributed cluster in the sensor networks. Finally the accurate data are transmitted to the sink node by each distributed cluster in the sensor network.

### *A. Data Correlation Among Sensor Nodes in Spatial Domain*

We are interested to discuss a mathematical model [16] to find the spatial data correlation among sensor nodes $i$ and $j$ to measure a tracing point [ 4]. Tracing point is a reference value or an event which we are going to measure in the sensor field (spatial domain). For example, we are interested to measure the moisture content in various location of agricultural field. According to literature [3] as the sensor node density increases, the spatial correlation between observed data $(S_i, S_j)$ among sensor nodes $i$ and $j$ also increases in the sensor region. The sensor nodes $i$ as well as $j$ can sense and measure the tracing point to collect the continuous data samples $S_i=\{ s_{i1} , s_{i2}, s_{i3}, \ldots\ldots s_{in} \}$ and $S_j=\{s_{j1} , s_{j2}, s_{j3}, \ldots\ldots s_{jn}\}$ over a window frame of time interval $T$.

We compute the mean, variance and covariance [16] for continuous data sample over a window frame of time interval $T$ for the sensor nodes $i$ and $j$ . Finally we find the correlation coefficient $(\rho_{S_i S_j})$ between observed data $(S_i, S_j)$ among sensor nodes $i$ and $j$ given by

$$\rho_{S_iS_j} = \frac{Cov(S_iS_j)}{Var(S_i)Var(S_j)} \quad (1)$$

The spatial correlation between observed data $(S_i, S_j)$ among sensor nodes $i$ and $j$ can be modeled as Joint Gaussian Random Variables (JGRV) [12,14 ] as follows.

$E[S_i]=0$, $E[S_j]=0$ ; $Var[S_i]=\sigma_{S^i}^2$ , $Var[S_j]=\sigma_{S^j}^2$ ; $Cov[S_i,S_j]=\sigma_{S^i}^2 Corr[S_i,S_j]$ for $i=1,2,............n$ and $j=1,2,.....................n$.

We define a correlation model [14] $K_V(.)$ which is given as

$$K_V(d_{i,j}) = Corr[S_iS_j] = \rho[S_iS_j] = \frac{E[S_iS_j]}{\sigma_{S^i}^2} = \frac{Cov[S_iS_j]}{\sigma_{S^i}^2} \quad (2)$$

where $d_{i,j} = \| S_i - S_j \|$ is the Euclidian distance between the sensor nodes $i$ and $j$ for the sensed data. The covariance function is a non-negative function and can decrease monotonically with distance $d_{i,j} = \| S_i - S_j \|$ with limiting value of 1 at $d=0$ and of 0 at $d=\infty$. Usually covariance models can be classified into four groups : Spherical , Power Exponential , Rational quadratic , Matern [21,22] . Out of these, power exponential model is widely used because the physical phenomenon such as electromagnetic waves can be expressed as exponential autocorrelation function [31,32]. Hence in this paper, we adopt power exponential model for correlation model $K_V(.)$ given as

$$K_V^{PE}(d) = e^{(-d/\theta_1)^{\theta_2}} \text{ for } \theta_1 > 0, \theta_2 \in (0,2] \quad (3)$$

We define a threshold $\alpha$ for $0 < \alpha \le 1$ which follows two properties:

- If $\rho_{S_iS_j} \ge \alpha_{ij}$ , observed data are strongly correlated among sensor nodes $i$ and $j$ in the sensor field.
- If $\rho_{S_iS_j} < \alpha_{ij}$, observed data are weakly correlated among sensor nodes $i$ and $j$ in the sensor field.

From (1),(2) and (3) , we formulate the correlation coefficient $\rho_{S_iS_j}$ for observed data among sensor nodes $i$ and $j$ using power exponential model under assumption that the data are strongly correlated in the spatial domain given as follows:

$$\rho_{S_iS_j} = \frac{Cov[S_iS_j]}{\sigma_{S_i}^2} = e^{(-d_{ij}/\theta_1)^{\theta_2}} \ge \alpha$$

(4)

From (4), we find the relation between the threshold value $\alpha$ and power exponential model given as

$$e^{(-d_{ij}/\theta_1)^{\theta_2}} \ge \alpha$$

$$d_{ij}^2 \le \theta_1^2 \sqrt[\theta_2]{\left(\log\left(\frac{1}{\alpha}\right)\right)^2} \quad (5)$$

We compare (5) with the Euclidean distance among the coordinates of sensor nodes $i$ and $j$ as follows.

$$d_{ij}^2 = (x_i - x_j)^2 + (y_i - y_j)^2 \quad (6)$$

From (5) and (6), we express

$$(x_i - x_j)^2 + (y_i - y_j)^2 \le \theta_1^2 \sqrt[\theta_2]{\left(\log\left(\frac{1}{\alpha}\right)\right)^2} \quad (7)$$

Comparing (7) with the equation of a circle , we get

$$(x_i - x_j)^2 + (y_i - y_j)^2 = R^2 \quad (8)$$

From *(7) and (8)*, we get the radius $R$ of circular data correlation range of each sensor node $i$ located in the sensor field represented as a centre coordinate .

$$R^2 \le \theta_1^2 \sqrt[\theta_2]{\left(\log\left(\frac{1}{\alpha}\right)\right)^2} \quad (9)$$

The neighboring sensor nodes $j$ which falls under the circular data correlation area range of each sensor nodes $i$, the observed data among sensor nodes $i$ and $j$ are highly correlated in the sensor field. Radius $R$ of data correlation range is implemented to form distributed clusters in the sensor field which is discussed bellow.

*B. Distributed Clustering Algorithm*

We form distributed non-overlapping clusters of irregular shape and size among the sensor nodes in the sensor field using a distributed clustering algorithm [16]. We construct the distributed clustering algorithm as follows:

*Notations used in the algorithm:*

$M$ =Set of sensor nodes deployed in the field

$\mathbb{C}$ =Set of cluster where each element $c \in \mathbb{C}$ is of the form $c=(a_c, b_c)$ where $a$ denotes the $CH$ node of the cluster $c$ and $b$ denotes the associated nodes (non-CH nodes) of the cluster $c$.

$R$ = Radius of data correlation range.

**Distributed Clustering Algorithm**

1. Start
2. Set $W = M$
3. $\forall i \in W$ ,let $G(i) = \{ j \in W : d(i,j) \le R, i \ne j \}$
   Where $d(i,j)$ is the Euclidean distance between $i$ and $j$ sensor nodes.
4. $S = \{ j \in M : G(j) = \max G(i) \}$ , $i \in M$ , we define $d_{\max}(i) = \max_{j \in G(i)} d(i,j)$ ,where $d(i,j)$ is the Euclidian distance between nodes $i$ and $j$.
5. Let $K = \arg\min_{i \in S} d_{\max}(i)$
6. $\mathbb{C} = \mathbb{C} \cup \{(K, G(K)\}$
7. $W = W - \{K\} - G(K)$
8. If $W \ne \{\varnothing\}$ , go to step 3

9. Stop

Finally non-overlapping distributed clusters are formed in the sensor field. Now we are interested to perform the data accuracy estimation for each distributed cluster in the sensor field which we explain in the next part.

*C. Clustering Algorithm based Data Accuracy Estimation*

As distributed non-overlapping clusters are formed in the sensor field using the clustering algorithm, we perform data accuracy [16] estimation for each distributed cluster. Each distributed cluster can measure a single tracing point of the same event in sensor field. We calculate the data accuracy for the measured data at the CH node of the respective cluster in the sensor field. The data accuracy is performed to check the estimated data received at the CH node from all the sensor nodes in the cluster are accurate and don't contain any redundant information.

We discuss the mathematical expression of data accuracy estimation for each distributed cluster with $M$ different sensor nodes in the field. Each sensor node $i$ can observe sense and measure the physical phenomenon of data $S_i$ for the tracing point value $S$ with observation noise $n_i$ for each distributed cluster. Therefore the observation done by the sensor node $i$ in each distributed cluster is given as

$$x_i = s_i + n_i \quad \text{where } i \in M \tag{10}$$

Each sensor node $i$ sense the observed sample data $x_i$ and transmits $x_i$ to the CH node in each distributed cluster sharing wireless Additive White Gaussian Noise(AWGN) channel [12,23 ] where $n_i$ is independent of each other and modeled as Gaussian Random Variable of zero mean and variance $\sigma_n^2$. Thus the observed sample data $x_i$ passes through AWGN channel to the CH node for each distributed cluster in the sensor field which reconstructs estimation $\hat{S}$ of the tracing point $S$. The CH node receive all $M$ observation sample for each distributed cluster given by

$$X = AZ + N \tag{11}$$

where $X$ is a $M \times 1$ data vector for observation done by $M$ different sensor nodes in each distributed cluster , $Z$ is a $(1+M)\times 1$ random vector for physically sensed data $S_i$ for $i \in M$ including the point event $S$ where we estimate for $\mathcal{N}(0, C_Z)$ , A is a known $M \times (1+M)$ matrix and $N$ is a $M \times 1$ noise vector for the observed data of $M$ different sensor nodes with $\mathcal{N}(0, C_Z)$ in each distributed cluster in sensor filed. The random vector Z with zero mean and covariance $\mathcal{N}(0, C_Z)$ can be shown as follows :

$$\mu_z = E[Z] = \begin{bmatrix} E[s] \\ E[s_1] \\ E[s_2] \\ . \\ . \\ E[s_M] \end{bmatrix} = \begin{bmatrix} 0 \\ 0 \\ 0 \\ . \\ . \\ 0 \end{bmatrix} \quad \text{and}$$

$$C_z = E[ZZ^T] = \begin{bmatrix} E[s,s] & E[ss_1] & E[ss_2] & . \; . & E[ss_M] \\ E[s_1 s] & E[s_1 s_1] & E[s_1 s_2] & . \; . & E[s_1 s_M] \\ E[s_2 s] & E[s_2 s_1] & E[s_2 s_2] & . \; . & E[s_2 s_M] \\ . & . & . & . \; . & . \\ . & . & . & . \; . & . \\ E[s_M s] & . & . & . \; . & E[s_M s_M] \end{bmatrix}$$

Thus the covariance matrix is

$$C_z = \sigma_s^2 \begin{bmatrix} 1_{1\times 1} & r_{1\times M}^T \\ r_{M\times 1} & B_{M\times M} \end{bmatrix} \tag{12}$$

where

$$r_{M\times 1} = \begin{bmatrix} \rho_{s_1,s} \\ \rho_{s_2,s} \\ . \\ . \\ \rho_{s_M,s} \end{bmatrix}$$

$$B_{M\times M} = \rho_{s_i,s_j} = \begin{bmatrix} \rho_{s_1,s_1} & \rho_{s_1,s_2} & \cdot & \rho_{s_1,s_M} \\ \rho_{s_2,s_1} & \rho_{s_2,s_2} & \cdot & \rho_{s_2,s_M} \\ \cdot & \cdot & \cdot & \cdot \\ \rho_{s_M,s_1} & \rho_{s_M,s_2} & \cdot & \rho_{s_M,s_M} \end{bmatrix}$$

In the matrix $C_Z$ , $r_{M\times 1} = \rho_{S_i,S}$ illustrates the correlation coefficient between $S_i$ , $S$ respectively and $B_{M\times M} = \rho_{S_i,S_j}$ is the correlation coefficient between $S_i$ , $S_j$ respectively. The power exponential model [21,22] is used for the correlation model to show the relation between $S_i$ and $S$ as well as $S_i$ and $S_j$. Thus we get $\rho_{S_i,S} = e^{-(d_{S,i}/\theta_1)^{\theta_2}}$ and $\rho_{S_i S_j} = e^{-(d_{i,j}/\theta_1)^{\theta_2}}$ in the covariance matrix $C_Z$ .The CH node collects all the observations from $M$ different sensor nodes in each distributed cluster to calculate the estimate of $\hat{S}$ from $\hat{S}_i$ . If the observed data $X$ can be modeled by Bayesian Linear Model [24 ] for all sensor nodes in each distributed cluster , the MMSE estimator to estimate the tracing point at the CH node in each cluster is given as :

$$\hat{Z} = E\left[Z \mid X\right] = \begin{bmatrix} \hat{s} \\ \hat{s}_1 \\ \hat{s}_2 \\ . \\ \hat{s}_M \end{bmatrix}$$

$$\hat{Z} = C_Z A^T (A C_Z A^T + \sigma_N^2 I_{M\times M})^{-1} X$$

$$\hat{Z} = \begin{pmatrix} r^T \\ B \end{pmatrix} \left( B + \frac{\sigma_N^2}{\sigma_S^2} I_{M\times M} \right)^{-1} X \qquad (13)$$

*Note : $\hat{Z} = E\left[Z \mid X\right]$ can be represented in MMSE as well as in MAP estimator as follows :*

*For MMSE : $\hat{Z} = \arg\min_{Z} E\left[\sum_{i=1}^{M} (Z_i - X_i)^2\right] = E\left[Z \mid X\right]$*

*For MAP : $\hat{Z} = \arg\min_{Z} P\left(Z \mid X\right) = E\left[Z \mid X\right]$*

*Both MMSE and MAP have the same meaning for $E\left[Z \mid X\right]$ in $\hat{Z}$ . We use MMSE estimator in this paper to find the data accuracy for each distributed cluster in the network.*

The measurement for the MMSE estimator at the CH node for each distributed cluster is given as the error $\in = (S - \hat{S})$ with mean zero and covariance matrix illustrated as

$$E[(Z - \hat{Z})(Z - \hat{Z})^T]$$

$$= C_Z - C_Z A^T (A C_Z A^T + \sigma_N^2 I_{M\times M})^{-1} A C_Z$$

$$= \sigma_S^2 \begin{pmatrix} 1 & r^T \\ r & B \end{pmatrix} - \sigma_S^2 \begin{pmatrix} r^T \\ B \end{pmatrix} \left( B + \frac{\sigma_N^2}{\sigma_S^2} I_{M\times M} \right)^{-1} (r\ B) \qquad (14)$$

From (14), we calculate the estimation of tracing point $(\hat{S})$ at the CH node in each distributed cluster in the sensor region given as

$$\hat{S} = r^T \left( B + \frac{\sigma_N^2}{\sigma_S^2} I_{M\times M} \right)^{-1} X \qquad (15)$$

We calculate the distortion factor between $S$ and $\hat{S}$ to perform data accuracy estimation at the CH node for each distributed cluster in the sensor field. From equation no (14), we get the distortion factor as

$$\bar{D} = E[(S - \bar{S})^2]$$

$$\bar{D} = \sigma_S^2 - \sigma_S^2 r^T \left( B + \frac{\sigma_N^2}{\sigma_S^2} I_{M\times M} \right)^{-1} r \qquad (16)$$

We normalize the distortion factor and finally calculated the data accuracy estimation of $M$ different sensor nodes for each distributed cluster in the sensor field is given as

$$D_A(M) = 1 - \frac{\bar{D}}{\sigma_S^2}$$

$$D_A(M) = r^T \left( B + \frac{\sigma_N^2}{\sigma_S^2} I_{M\times M} \right)^{-1} r \quad \text{where} \quad \left( B + \frac{\sigma_N^2}{\sigma_S^2} I_{M\times M} \right)^{-1} r = \beta$$

$$D_A(M) = r^T \beta \qquad (17)$$

Data accuracy $D_A(M)$ for estimations is defined in terms of expectation of the errors between the actual point event and mean square average estimate of $M$ sensor nodes. $D_A(M)$ calculated at the CH node for each distributed cluster in the sensor field is performed and finally transmits the most appropriate data to the sink node. Hence the purpose of verifying the data accuracy $D_A(M)$ at CH node for each distributed cluster is to confirm that the most accurate data transmitted by $M$ different sensor node can aggregate rather than aggregating all the redundant data at the CH node [16]. The information accuracy proposed in literature [12] shows that at first each sensor nodes $i$ can calculate the MMSE estimate $\hat{S}_i$ for observed data and then transmits the estimated data $\hat{S}_i$ to the CH node i.e. $\hat{S}_i$ in order to find $\hat{S}$ . Finally averaging all $\hat{S}_i$ at the CH node for the cluster for $M$ sensor nodes to get $\hat{S}$ .But in Data accuracy estimation $D_A(M)$ , at first we collect all the observed data from $M$ sensor nodes and then only perform the MMSE estimation at the CH node for each distributed cluster. It is efficient to calculate the MMSE estimation only at the CH node rather than performing the MMSE estimation at individual nodes and then averaging it at the CH node for each distributed cluster in the sensor field.

| Cluster Number | Cluster Head Node ID | Associated Nodes ID in Cluster | Information Accuracy I(M) | Data Accuracy D(M) |
|---|---|---|---|---|
| I | 14 | 12,16,17,23,24,26,28,29 | 0.8502 | 0.9577 |
| II | 1 | 5,7,11,13,15,18,19 | 0.8690 | 0.9632 |
| III | 10 | 2,3,8,20,21,27,30 | 0.8769 | 0.9585 |
| IV | 9 | 4,25 | 0.8732 | 0.9459 |
| V | 6 | 22 | 0.8349 | 0.9627 |

Table 1: Distributed clusters with data accuracy estimation

According to the clustering algorithm discussed in Section II B with threshold value $\alpha$ =0.7, we form five non overlapping distributed clusters of different sizes after performing simulation as shown in Table-1. Each distributed cluster has its CH node with associated sensor nodes which perform the data accuracy at the CH node. We compare the degree of accuracy for information accuracy model [12] with our data accuracy $D_A(M)$ estimation which shows $D_A(M)$ always perform better than $I(M)$ for each distributed cluster in the network.

We find that each distributed cluster in the network perform data accuracy $D_A(M)$ at their respective CH node and transmits the most accurate data to the sink node. We take the first cluster having nine sensor nodes given in Table-I which we get after simulating the clustering algorithm. Fig.1, shows that six sensor nodes are sufficient to perform the same data accuracy level achieve by nine sensor nodes in the cluster with $\theta_1 = 70$ and $\theta_2 = 1$ . Hence, we only choose certain number of sensor nodes which are close to the tracing point instead of choosing all sensor nodes in each distributed cluster subjected to achieve approximately good data accuracy in the network. Thus we can reduce the number of sensor nodes in each distributed cluster subjected to data accuracy estimation in the sensor network.

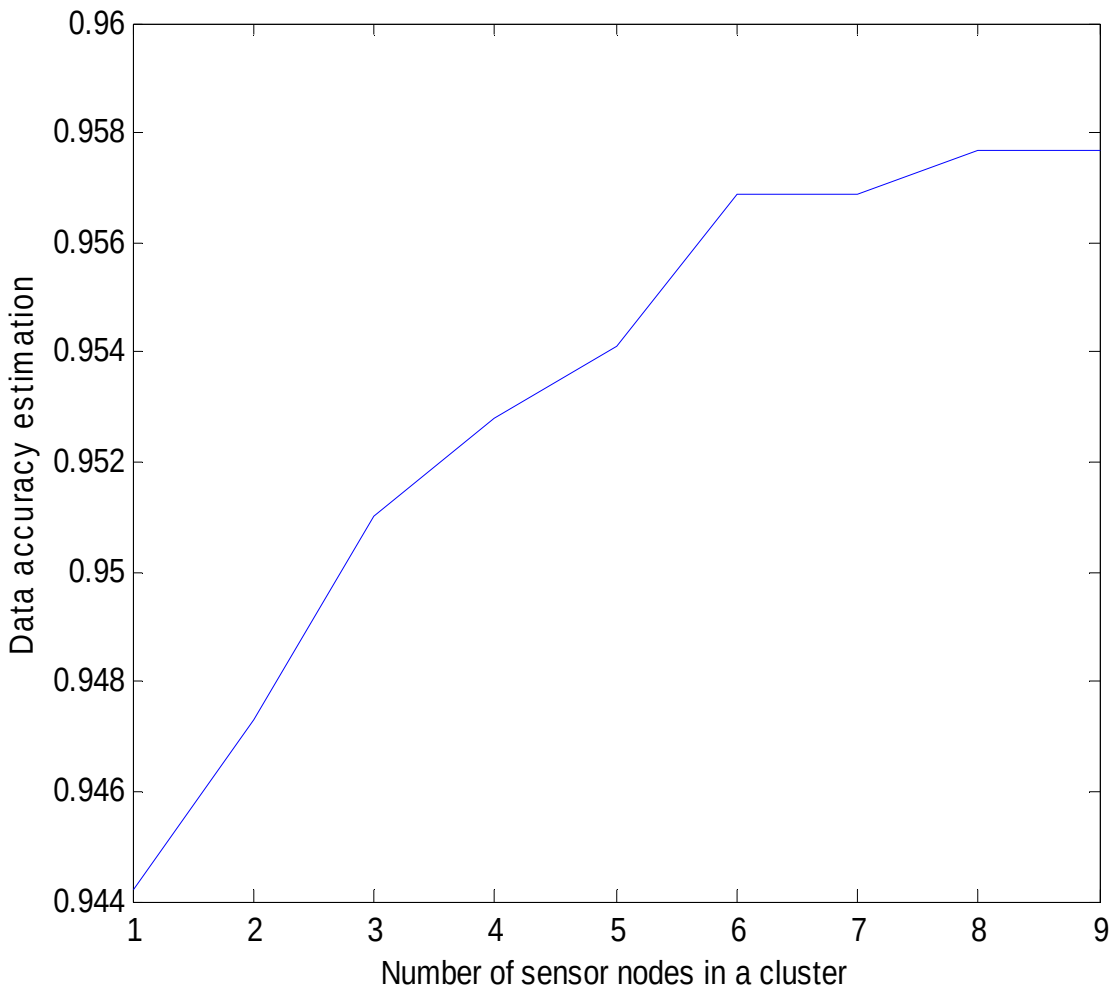

Figure 1: Number of sensor nodes vs. data accuracy in a cluster

## III. PROBABILISTIC MODEL FOR NODE SELECTION IN THE NETWORK

In the previous section, we find the data accuracy for each distributed cluster and conclude that there exists an optimal number of sensor nodes in each distributed cluster. Moreover these optimal sensor nodes in each distributed cluster are sufficient to perform approximately the same data accuracy achieve by each cluster. Hence it is unnecessary to choose all the sensor nodes in each distributed cluster to perform the transmission process in the network. Optimal number of sensor nodes which are active can perform the data accuracy and rest of the sensor nodes goes to sleep mode in each distributed cluster in the network. Since an optimal number of sensor nodes are active and rests of them are in sleep mode in each distributed cluster, we perform an analytical as well as simulation model for energy consumption in the networks to save energy. Each distributed cluster is independent to each other irrespective of transmission of data packets to the sink node. Each distributed cluster has its own intelligence to operate its associated sensor nodes to switch on (active mode) or switch off (sleep mode). Since each distributed cluster has its own control to operate the associate sensor nodes, we find the probability by which the sensor nodes are in active mode and sleep mode subjected to data accuracy and energy consumption.

### *A. Probability by which Sensor Nodes are Active Subjected to Data Accuracy*

Here we perform analytical and simulation model for the sensor nodes which are active subjected to data accuracy in each distributed cluster in the network. We demonstrate a probabilistic model for sensor nodes which are in active mode and rest of the sensor nodes are in sleep mode in each distributed cluster in the network. Since in each distributed cluster in the network, CH node performs the data accuracy for associated nodes in the cluster, CH node is always switched on. The associated nodes are in active mode with probability *P* and in sleep mode with probability *(1-P)* in each distributed cluster in the network. We go through the following steps to validate our model:

M= Total number of sensor nodes in each distributed cluster in the network
N= Total number of sensor nodes which are switch on (active mode) in each distributed cluster in the network
$X_0$ =CH node always switches on in each distributed cluster.
$m*$=Expected number of sensor nodes which are in active mode in each distributed network

Total number of sensor nodes which are in active mode for each distributed cluster in the network are given as

$$N = X_0 + \overline{X}$$
$$N = X_0 + \sum_{i=1}^{M-1} X_i \tag{18}$$

Where

$X_i$ = 1, if nodes $i$ selected with probability *P*, (sensor nodes with switch on)

$X_i$ =0, if nodes $i$ selected with probability *(1-P)*, (sensor nodes with switch off)

Expected number of sensor nodes which are in active mode (switch on) with probability *P* in each distributed cluster in the network is given as

$$E(N) = E(X_0) + \sum_{i=1}^{M-1} E(X_i)$$
$$E(N) = 1 + (M-1)P$$
$$E(N) = MP - P + 1$$
$$E[N] = m* \tag{19}$$

Probability by which the sensor nodes are in active mode (switch on) in each distributed cluster in the network is given as

$$P = \frac{m*-1}{M-1} \tag{20}$$

We find all the combinations of $M$ sensor nodes taken $K$ at a time in each distributed cluster which are given as ${}^{M}C_{K}$ . Since the CH node is fixed for every combination in each distributed cluster in the network, the combination becomes ${}^{M-1}C_{K-1}$ for the number of all the combinations of $M$ sensor nodes taken $K$ at a time given as.

$$^{M}C_{K} = \frac{M!}{K!(M-K)!}$$

$$= \frac{M(M-1)!}{K(K-1)!((M-1)-(k-1))!}$$

$$= \frac{M}{K}\, {}^{M-1}C_{K-1}$$

or $\quad {}^{M-1}C_{K-1} = \frac{K}{M} C_K^M \qquad (21)$

*Proof that:* $\quad {}^{M-1}C_{K-1} = \frac{K}{M} C_K^M$

$$^{M-1}C_{K-1} = \frac{M(M-1)!}{M(K-1)!((M-1)-(K-1))!}$$

$$= \frac{M!}{M(K-1)!(M-K)!}$$

$$= \frac{(M-K+1)M!}{M(M-K+1)(K-1)!(M-K)!}$$

$$= \frac{(M-K+1)}{M}\left[\frac{M!}{(K-1)!(M-K+1)!}\right]$$

$$= \frac{(M-K+1)}{M}\left[\frac{M!}{(K-1)!(M-K+1)(M-K)!}\right]$$

$$= \frac{K}{M}\left[\frac{M!}{K(K-1)!(M-K)!}\right]$$

$$= \frac{K}{M}\left[\frac{M!}{K!(M-K)!}\right]$$

$$= \frac{K}{M}\, {}^{M}C_{K}$$

The probability mass function (pmf) for active nodes in each distributed cluster with a fixed CH node in the network can be represented by binomial distribution given as.

$$P(N=K) = P(\overline{X} = K-1)$$

$$= {}^{M-1}C_{K-1} P^{K-1}(1-P)^{M-K}$$

$$= \frac{K}{M}\, {}^{M}C_{K} P^{K-1}(1-P)^{M-K} \qquad (22)$$

The probability of sensor nodes which are in active mode subjected to data accuracy function $A(P)$ in each distributed cluster in the network is given by

$$A(P) = \sum_{K=1}^{M} \frac{K}{M} A(N=K)\, {}^{M}C_{K} P^{K-1}(1-P)^{M-K} \qquad (23)$$

Now we find a minimum probability ($P_{\min}$) with $\chi_a$ as a user dependent factor $(0 < \chi_a < 1)$ to switch on the sensor nodes in the network. Thus the minimum probability for achieving a required level of data accuracy to switch on the sensor nodes in each distributed clusters is given as

$$P_{\min} = \arg\min_{P}\{A(P) \geq \chi_a A_{\max}\} \quad \text{where } A_{\max} = A(1) \qquad (24)$$

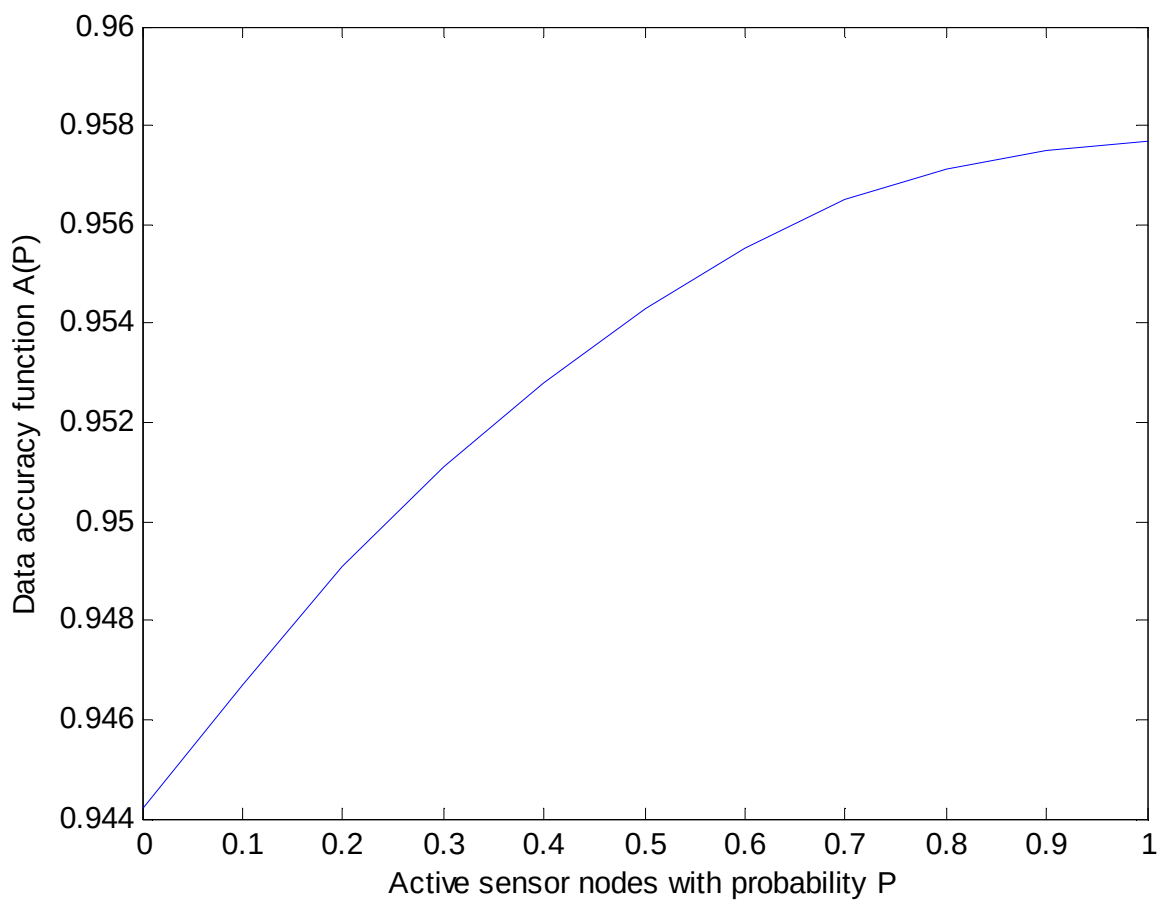

Figure 2: Data accuracy function vs active sensor nodes with probability P

By implementing the clustering algorithm with threshold value $\alpha$ =0.7 discussed in *section II (B),* we form distributed non overlapping clusters with different sizes in the network. We choose the first cluster with nine sensor nodes from the network to satisfy *(23),* which performs the data accuracy for this cluster as shown in Fig. 2. We illustrate the analysis of this cluster in probabilistic form subjected to data accuracy function. In Fig. 2, we plot the probability of sensor nodes which are in active mode for the clusters with respect to data accuracy. We take minimum probability $P_{\min}$=0.7 for achieving a required level of data accuracy $A(P)$ for $\chi_a$ =0.956 to switch on the sensor nodes in the cluster. So with $P_{\min}$ a certain number of sensor nodes are switched on which are sufficient to give approximately the same data accuracy level achieve by the all the sensor nodes in each distributed cluster. Thus each an every distributed cluster can switch on certain number of sensor nodes (generally sensor nodes close to tracing point) with minimum probability $P_{\min}$ and rest of them goes to sleep mode in each distributed cluster in the network. Thus an optimal number of sensor nodes in each distributed cluster are sufficient to operate with $P_{\min}$ in the networks respect to data accuracy instead of switch on all the sensor nodes in the networks. Hence we can save node deployment cost per cluster by operating $P_{\min}$ in different distributed cluster in the network.

### *B. Probability by which Sensor Nodes are Active Subjected to Energy Consumption*

Each distributed cluster can perform the energy consumption model [6] in the network. We analyze and simulate a probabilistic model for energy consumption of each distributed cluster in the network. For our simplicity we model the energy consumption for each distributed cluster and then implement it to calculate the energy consumption for the whole network. For each distributed cluster in the network, energy consumption is given as

$$E_{Cluster} = E_{CH} + E_{Non-CH}(M-1) \qquad (25)$$

where $E_{CH}$ is the energy consumed for the $CH$ node, the energy consumed for non-CH nodes is given by $E_{Non-CH}$ and $M$ is the total number of sensor nodes in each distributed cluster.

We take a radio hardware energy dissipation model [6] ,in which transmitter antenna dissipates energy for the radio electronics , power amplifier and the receiver antenna dissipates energy to run the radio electronics . We use both the free space ($d^2$ power loss) channel model and the multipath fading ($d^4$ power loss) channel model which is dependent on the distance between transmitter and receiver antenna respectively [30].

We define a threshold $\tau_0$ ,if the distance is less than the threshold $\tau_0$, the free space model is used otherwise we can use multipath model [6]. Hence the energy consumption to transmit the data is given as

$$\begin{aligned} E_{TX} &= lE_{Elec} + l\in_{fs} d^2 && \text{for } d < \tau_0 \\ &= lE_{Elec} + l\in_{mp} d^4 && \text{for } d \ge \tau_0 \end{aligned} \qquad (26)$$

Energy consumed to received the data is given as $E_{RX} = lE_{Elec}$. $E_{Elec}$ is a electronics energy which depends upon spreading of signal , modulation techniques and coding to be used. $\in_{fs} d^2$ and $\in_{mp} d^4$ are the amplifier energy depends upon the distance between the transmitter and the receiver antenna respectively . The parameters for communication energy are given as: $E_{Elec}$ =*50nJ/bit*, $\in_{fs}$ =*10pJ/bit/m*$^2$, $\in_{mp}$ =0.0013pJ /bit/$m^4$ and the data aggregation is set as $E_{Agg}$ =5nJ/bit /signal.

Each non-CH nodes in a cluster only transmits the data to the CH node. Moreover the distance between the non-CH nodes to the CH node is small in each distributed cluster in the network. Hence we adopt free space channel model ($d^2$ power loss).Thus the energy consumption in each non-CH node in a cluster is given as

$$\begin{aligned} E_{Non-CH} &= E_{Tx} \\ &= lE_{Elec} + l\in_{fs} d^2_{toCH} \end{aligned} \qquad (27)$$

We define the sensor nodes (non-CH nodes) at a polar coordinate $(r,\theta)$ from the CH node in a cluster and integrate over the disk as follows.

$$E_{non-CH} = lE_{Elec} + l\in_{fs} \int_0^{2\pi}\int_0^{R} r^2 \rho r dr d\theta$$

Where $R$ is the radius of the correlation coefficient of a node discussed in Section II A and $\rho = \frac{M}{\pi R^2}$.

$$E_{non-CH} = lE_{Elec} + l\in_{fs}\left(\frac{MR^2}{2}\right) \qquad (28)$$

$M$ and $R$ varies in each distributed cluster in the network depending upon the cluster size. The energy consumption at the CH node for each distributed cluster can be given as receiving the data from the non-CH nodes, aggregating the data and transmitting the aggregate data to the base station. Since the base station is far away from each distributed cluster, we use the multipath channel model ($d^4$ power loss). Thus the energy consumed at the CH node in each distributed cluster is given as:

$$\begin{aligned} E_{CH} &= E_{RX} + E_{Agg} + E_{TX} \\ &= lE_{Elec}(M-1) + lE_{Agg}M + lE_{Elec} + l\in_{mp} d^4_{toBS} \end{aligned} \qquad (29)$$

From (28) and (29), the total energy consumption in each distributed cluster in the network is given as

$$E_{cluster} = E_{CH} + (M-1)E_{non-CH} \qquad (30)$$

Now we develop a probabilistic model for active sensor nodes in each distributed cluster subjected to energy consumption. The probability by which energy is consumed in each distributed cluster in the network is given as

$$\begin{aligned} E_{Cluster}(P) &= \sum_{K=1}^{M} E_{Cluster}(N=K)P(N=K) \\ &= \sum_{K=1}^{M} E_{Cluster}(N=K)\left[{}^{M-1}C_{K-1}P^{K-1}(1-P)^{M-K}\right] \\ &= \sum_{K=1}^{M} E_{Cluster}\left[\left(\frac{K}{M}C_K^M\right)P^{K-1}(1-P)^{M-K}\right] \end{aligned} \qquad (31)$$

Now we define a maximum probability $P_{\max}$ with $\chi_e$ a user dependent factor $(0<\chi_e<1)$ to switch on the sensor nodes in each distributed cluster subjected to energy consumption. Thus the maximum probability $P_{\max}$ for keeping the energy consumption below a certain level to switch on the sensor nodes in each distributed cluster in the network is given as

$$P_{\max} = \arg\max_P \{E_{cluster}(P) \le \chi_e E_{\max}\} \quad \text{where } E_{\max} = E_{cluster}(1) \qquad (32)$$

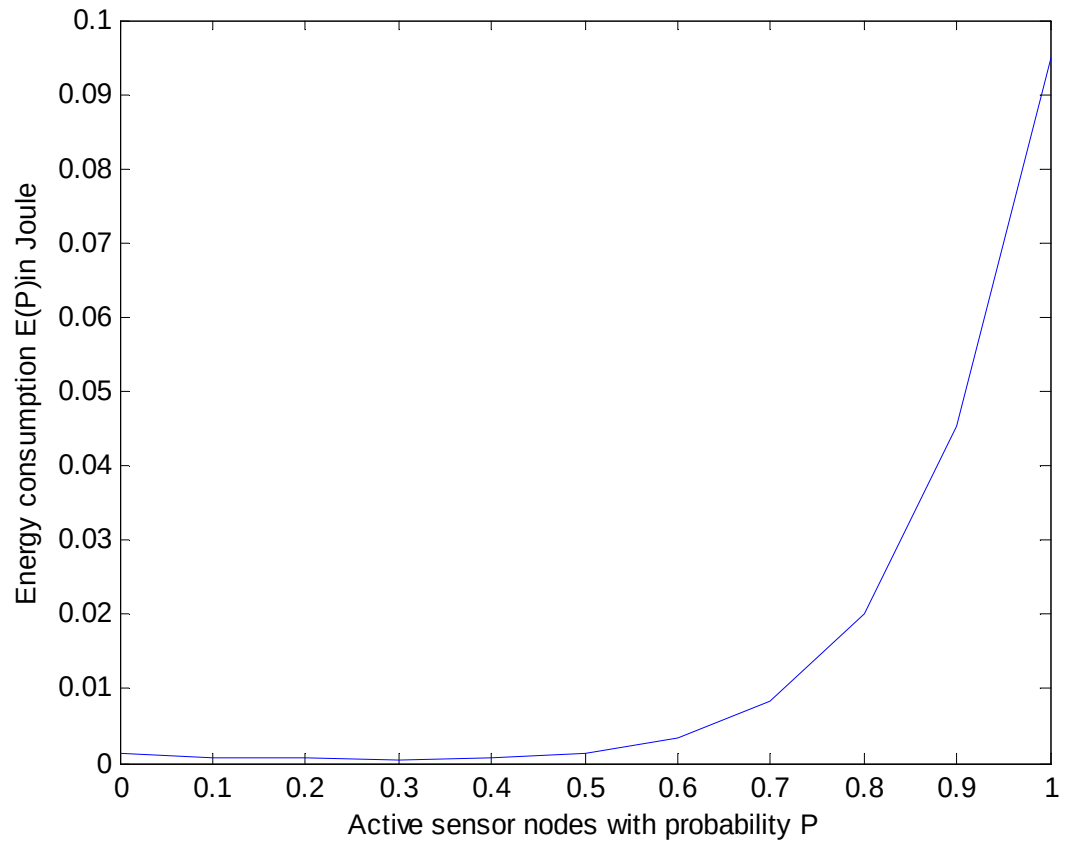

Figure 3: Energy consumption vs. active sensor nodes with probability P

We take the first cluster with nine sensor nodes chosen from the clustering algorithm with threshold value $\alpha$ =0.7 and perform the energy consumption model for the cluster. We analysis the energy consumption for the cluster in probabilistic form. In Fig. 3, we plot the probability of the sensor nodes which are switched on for the cluster of nine sensor nodes subjected to energy consumption. We take maximum probability $P_{\max}=0.6$ to keep the energy consumption $E_{cluster}(P)$ for $\chi_e = 0.004$ below a certain level to switch on the sensor nodes in the cluster. Thus it is clear from the Fig. 3 that we can operate $P_{\max}$ such that certain numbers of sensor nodes are switch on subjected to energy consumption rather than switching all the sensor nodes in each distributed cluster. Thus we save the energy cost by reducing the number of sensor nodes in each distributed clusters and finally optimized the energy consumption in the whole network by operating $P_{\max}$ .

*C. Trade Off Between Data Accuracy and Energy Consumption for Node Selection in the Network*

Previously we have illustrated that the sensor nodes are switched on and switched off in each distributed cluster in the network subjected to data accuracy and energy consumption in probabilistic approach. Thus it leads to the trade off between $P_{\min}$ for data accuracy and $P_{\max}$ for energy consumption for selecting an optimal sensor node in each distributed cluster in the network. Hence we define a probability $P$ such that $P \in [(P_{\min} + P_{\max})/2]$ to select an optimal number of sensor nodes which are sufficient to achieve the data accuracy $A(P)$ and can reduce the energy consumption $E_{cluster}(P)$ in each distributed cluster in the sensor network. The probability $P$ for the cluster of nine sensor nodes discussed in section III A-B is 0.65 and the expected number of sensor nodes selected in the first cluster are six out of nine sensor nodes. Hence we find an optimal expected number of sensor nodes by fixing an appropriate probabilistic value $P$ for active nodes in each distributed cluster having trade off between data accuracy and energy consumption. This leads to reduce the node deployment cost and energy cost per cluster in the distributed network.

| *#of clusters* | *Total No. of nodes in each cluster* | *Probability P* for selecting nodes* | *Expected No. of nodes E[N]=m* selected in each cluster* |
|---|---|---|---|
| *I* | *9* | *0.65* | *6* |
| *II* | *8* | *0.5* | *5* |
| *II* | *8* | *0.6* | *5* |
| *IV* | *3* | *0.4* | *2* |
| *V* | *2* | *0.5* | *2* |
| *Total no. of nodes in network* | *30 Nodes* | | *20 Nodes Selected in network* |

Table 2: Expected number of sensor nodes selected in the network

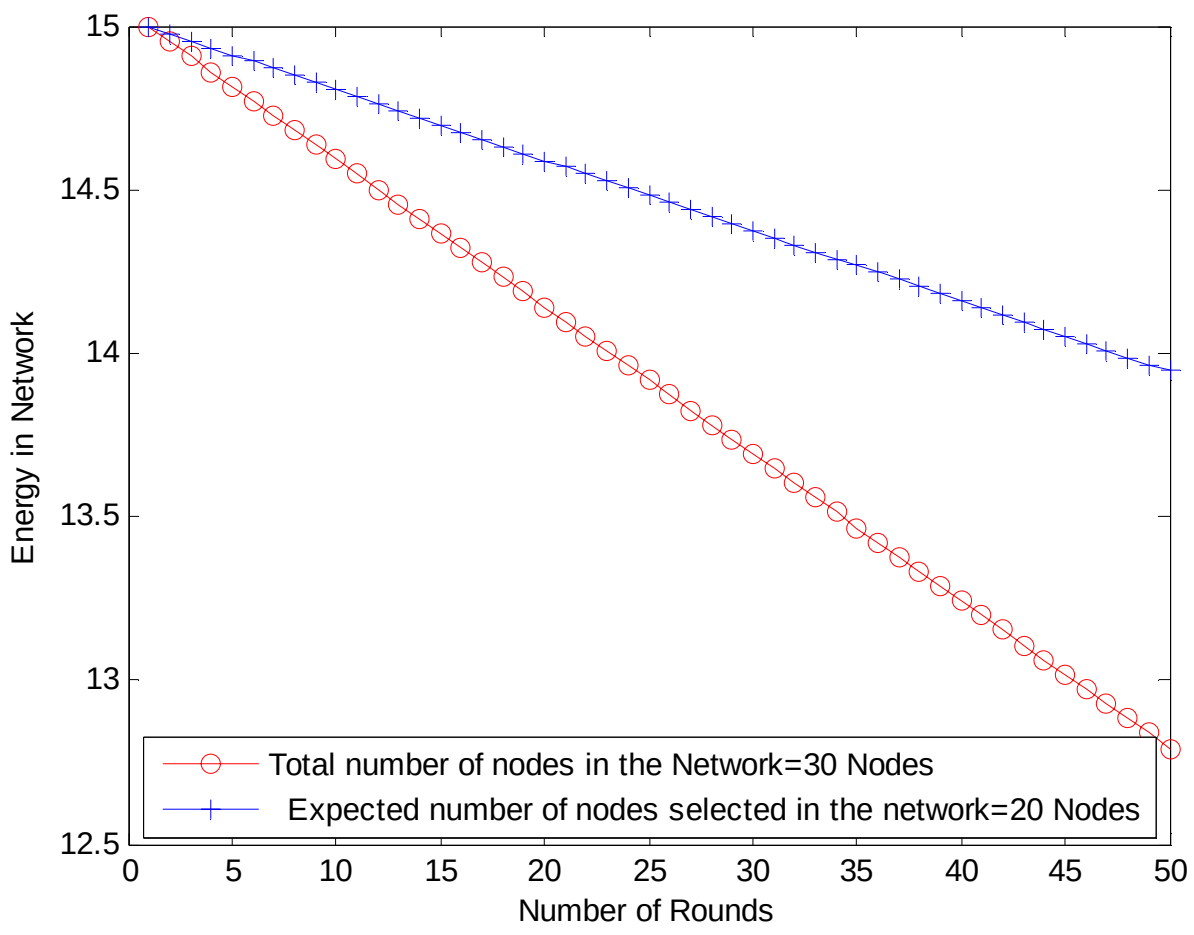

Figure 4: Energy in network vs. number of rounds

In Table-2, we are interested to show the expected number of sensor nodes selected in the network. Initially we deploy thirty sensor nodes in the sensor field. According to the clustering algorithm discussed in section-II B, sensor nodes form five distributed non-overlapping clusters in the sensor field. Each distributed cluster fixes a different probabilistic value $P$ to select active sensor nodes depending upon the trade off between data accuracy and energy consumption. Hence out of thirty sensor nodes, twenty sensor nodes are selected for data communication in the network. Finally we simulate energy consumption model for thirty sensor nodes and selected twenty sensor nodes as shown in Fig. 4. We conclude that energy consumption for thirty sensor nodes decays much faster than the selected twenty sensor nodes in the network with respect to time. Thus reducing the number of sensor nodes can save energy consumption in the whole network and increases life time of the network.

## IV. CONCLUSION

We conclude that we find the optimal number of sensor nodes using the trade off between data accuracy and energy consumption in probabilistic approach in each distributed cluster. This reduces the node deployment cost and energy cost in the network. We illustrate the probabilistic approach by which the sensor nodes can be operated for active mode and sleep mode using the trade off between data accuracy and energy consumption in each distributed cluster in the network. Simulation results shows energy consumption for the total number of sensor nodes deployed in the sensor region decays much faster than the optimal number of sensor nodes selected with respect to time. This increases life time of the network. Moreover our data accuracy estimation model performs better than information accuracy model in each distributed cluster with respect to data accuracy.

*Appendix-I: Summery for selecting optimal sensor nodes in each distributed cluster using trade off between data accuracy and energy consumption.*

| *#of clusters* | *No. of nodes in each cluster* | $P_{\min}$ *for Data accuracy* | $P_{\max}$ *for Energy consumption* | *Probability P for selecting nodes* | *Expected No. of nodes E[N]=m* selected* |
|---|---|---|---|---|---|
| I | 9 | 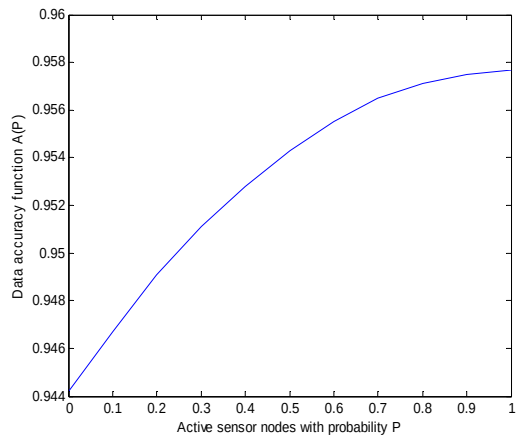<br>$P_{\min}$ *=0.7* | 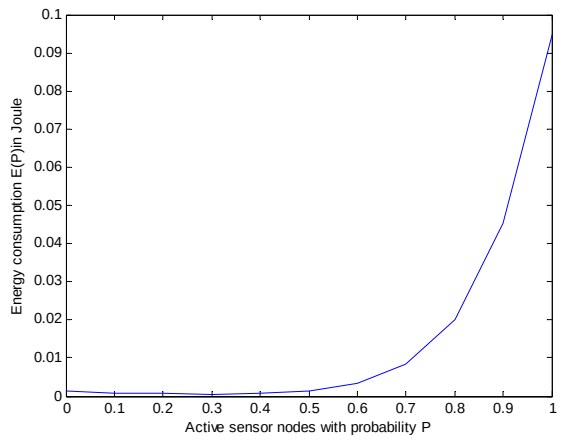<br>$P_{\max}$ *=0.6* | 0.65 | 6 |
| II | 8 | 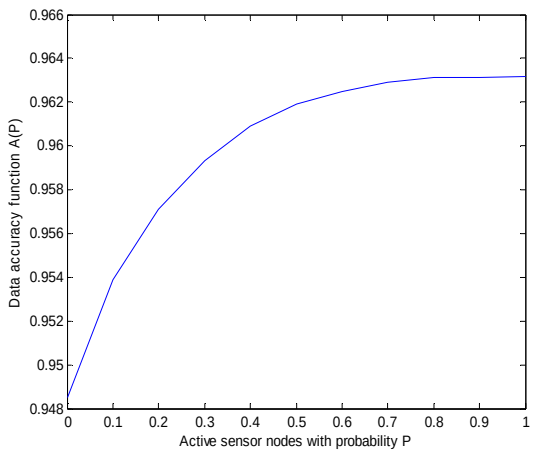<br>$P_{min}$ =0.5 | 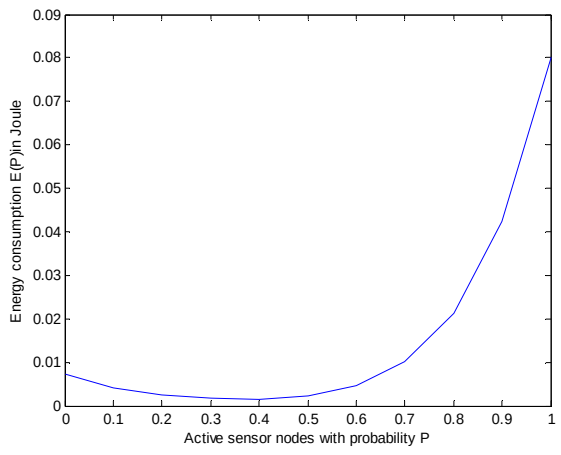<br>$P_{max}$ =0.5 | 0.5 | 5 |
| III | 8 | 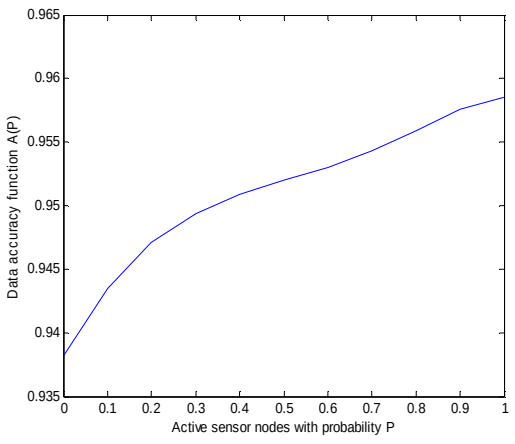<br>$P_{min}$ =0.8 | 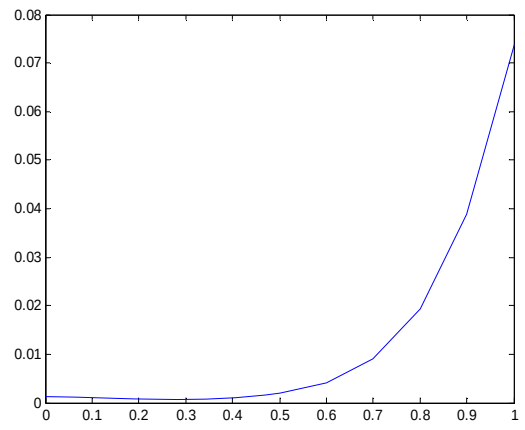<br>$P_{max}$ =0.4 | 0.6 | 5 |
| IV | 3 | 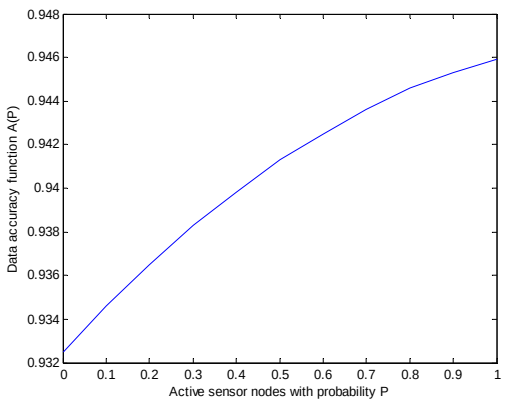<br>$P_{min}$ =0.5 | 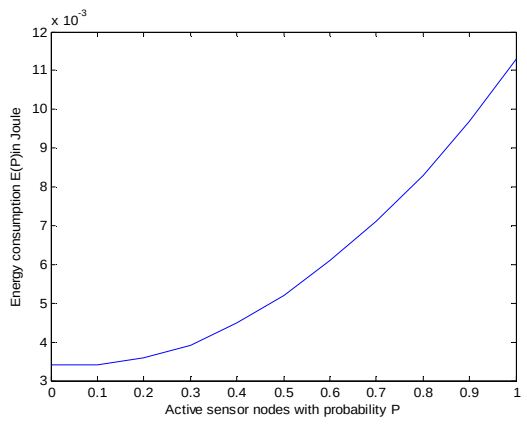<br>$P_{max}$ =0.3 | 0.4 | 2 |
| V | 2 | 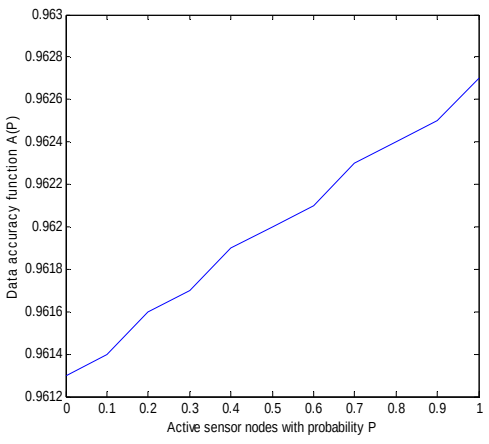<br>$P_{min}$ =0.5 | 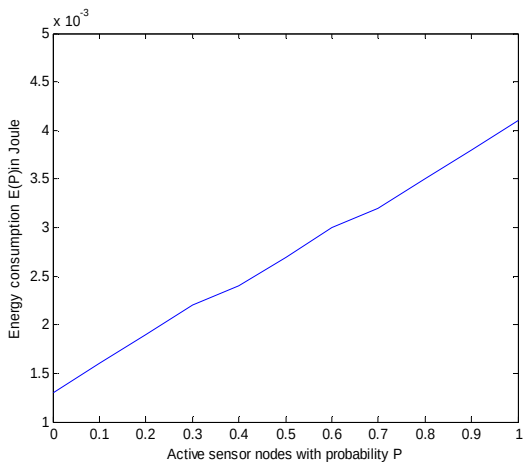<br>$P_{max}$ =0.5 | 0.5 | 2 |
| *Total nodes in N/W* | *30 Nodes* | | | | *20 Nodes selected* |